\documentclass[amsmath,aps,pra,reprint,superscriptaddress]{revtex4-2}

\pdfoutput=1
\usepackage[utf8]{inputenc}
\usepackage[T1]{fontenc}
\usepackage{lmodern}
\usepackage{microtype}

\usepackage{graphicx}
\usepackage{hyperref}
\usepackage{xcolor}

\definecolor{one}{HTML}{CC0000}

\hypersetup{
  colorlinks=true,
  citecolor=one,
  linkcolor=one,
  urlcolor=one
}

\newcommand*{\ev}[1]{\langle #1 \rangle}
\newcommand*{\bra}[1]{\langle #1 \rvert}
\newcommand*{\ket}[1]{\lvert #1 \rangle}

\begin{document}

\title{Error detection on quantum computers improves accuracy of
  chemical calculations}

\author{Miroslav Urbanek}

\email[Corresponding author: ]{urbanek@lbl.gov}

\affiliation{Computational Research Division, Lawrence Berkeley
  National Laboratory, Berkeley, CA 94720, USA}

\author{Benjamin Nachman}

\affiliation{Physics Division, Lawrence Berkeley National Laboratory,
  Berkeley, CA 94720, USA}

\author{Wibe A. de Jong}

\affiliation{Computational Research Division, Lawrence Berkeley
  National Laboratory, Berkeley, CA 94720, USA}

\begin{abstract}

A major milestone of quantum error correction is to achieve the
fault-tolerance threshold beyond which quantum computers can be made
arbitrarily accurate. This requires extraordinary resources and
engineering efforts. We show that even without achieving full fault
tolerance, quantum error detection is already useful on the current
generation of quantum hardware. We demonstrate this experimentally by
executing an end-to-end chemical calculation for the hydrogen molecule
encoded in the [[4, 2, 2]] quantum error-detecting code. The encoded
calculation with logical qubits significantly improves the accuracy of
the molecular ground-state energy.

\end{abstract}

\maketitle

\section{Introduction}

Quantum computing promises efficient methods for quantum-chemical
calculations that can reach far beyond the abilities of classical
computers~\cite{Reiher:2017}. Large-scale calculations will require an
ability to detect and correct errors. However, near-term devices known
as noisy intermediate-scale quantum (NISQ)
computers~\cite{Preskill:2018} are not expected to be fully
fault-tolerant. Despite this limitation, they can still be useful for
solving certain problems in physics and chemistry. In particular, the
variational quantum eigensolver (VQE)~\cite{Peruzzo:2014,
  McClean:2016} is an algorithm designed to work well on NISQ
computers. It has been experimentally demonstrated that VQE is able to
find the ground state as well as excited states of small quantum
systems encountered in quantum chemistry and nuclear
physics~\cite{Peruzzo:2014, OMalley:2016, Kandala:2017, Shen:2017,
  Colless:2018, Dumitrescu:2018, Hempel:2018, Ganzhorn:2019,
  Kokail:2019}. The performance of NISQ algorithms is currently
limited by gate errors and device noise. Several novel error
mitigation and suppression techniques have been developed to overcome
the imperfections of real devices~\cite{Li:2017, Temme:2017,
  McClean:2017, BonetMonroig:2018, Endo:2018, McArdle:2018, Endo:2019,
  Kandala:2019, McClean:2020, Otten:2019a, Otten:2019b,
  Sagastizabal:2019}.

Quantum error correction (QEC)~\cite{Gottesman:1997, Nielsen:2010,
  Devitt:2013, Terhal:2015, Campbell:2017} is a theory developed in
the last two decades to address this problem in a systematic way. An
important milestone for QEC experiments is to achieve the
fault-tolerance threshold. Fault tolerance requires a large number of
qubits, long coherence times, and low gate errors. However, QEC can
still be useful even without achieving fault tolerance and even with
only a small number of qubits~\cite{Gottesman:2016, Chao:2018a,
  Chao:2018b}. QEC can potentially increase coherence times and reduce
error rates in existing devices. There have been efforts to
demonstrate that quantum circuits using QEC codes can improve
accuracy, or at least break even, in comparison with the original
circuits. Previous experiments studied quantum codes that encode a
single logical qubit~\cite{Reed:2012, Nigg:2014} and also demonstrated
necessary improvements in qubit and gate qualities for
QEC~\cite{Barends:2014, Kelly:2015, Wootton:2018}. There has also been
a growing interest in studying the [[4, 2, 2]] quantum
code~\cite{Linke:2017, Takita:2017, Roffe:2018, Vuillot:2018,
  Willsch:2018}. Recently, it has been shown that logical gates
encoded in this code can achieve better fidelities than corresponding
physical gates~\cite{Harper:2019}. These efforts have tested
individual steps of QEC protocols separately. However, it has never
been demonstrated that an encoded calculation provides a tangible
benefit in practical applications.

In this work we demonstrate that QEC provides an improvement in
accuracy in an end-to-end quantum-chemical calculation. We have
implemented a two-qubit VQE algorithm for calculating the ground-state
energy of the hydrogen dimer in the [[4, 2, 2]] QEC
code~\cite{Vaidman:1996, Grassl:1997, Devitt:2013}. Instead of two
physical qubits, the calculation uses two logical qubits encoded in
four physical qubits. The code facilitates detection of a single
bit-flip and phase-flip error in either of the two logical qubits. Our
circuit additionally uses two ancillary qubits to perform a syndrome
measurement during the initial state preparation and to perform a
logical qubit rotation. Analytical simulations predict that the
encoded circuit should outperform the physical circuit up to a fairly
large error rate. We implement both the two-qubit and the six-qubit
circuit on the IBM Q Experience platform.

\section{Quantum algorithm}

Finding the ground-state energy of the $\mathrm{H}_2$ molecule in the
minimal basis is the simplest molecular electronic-structure
problem. It is often used as a benchmark and allows us to evaluate our
approach in comparison to earlier work~\cite{OMalley:2016,
  Kandala:2017, Colless:2018, Hempel:2018, Kandala:2019,
  Ganzhorn:2019}.

The $\mathrm{H}_2$ molecular Hamiltonian can be transformed into a
qubit Hamiltonian using the Jordan--Wigner~\cite{Jordan:1928},
Bravyi--Kitaev~\cite{Bravyi:2002}, or another similar
transformation. Here we use the explicit transformation defined in
Ref.~\cite{Colless:2018} that maps the subspace of the Hamiltonian
corresponding to two electrons with zero total spin to a two-qubit
Hamiltonian. The transformed Hamiltonian is given by
\begin{equation}
  \label{hamiltonian}
  H = g_1 + g_2 Z_1 + g_3 Z_2 + g_4 Z_1 Z_2 + g_5 X_1 X_2,
\end{equation}
where $X_i$, $Y_i$, and $Z_i$ denote Pauli operators acting on qubit
$i$ and $g_j$ are classically-calculated coefficients that depend on
the internuclear separation $R$. We use values of $g_j$ published in
Ref.~\cite{Colless:2018}.

\begin{figure*}
  \centering
  \includegraphics{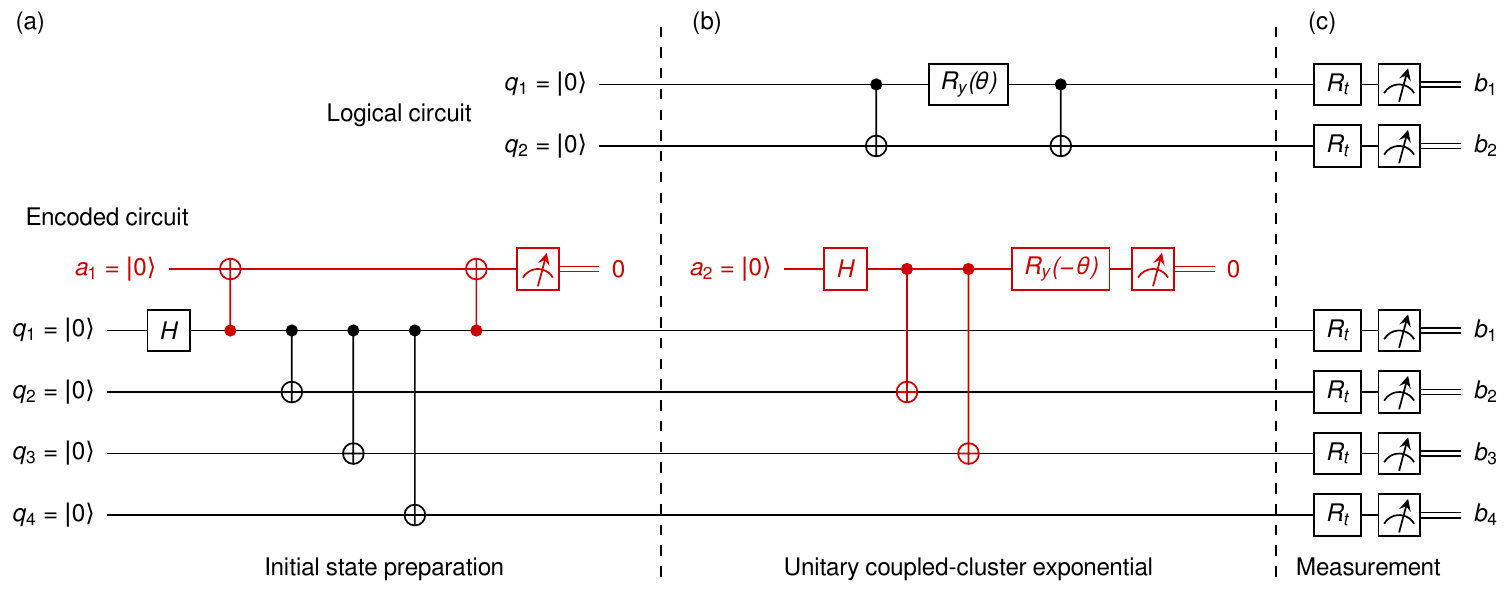}
  \caption{\label{circuits} Quantum circuits for the preparation of
    the UCC ansatz and for the measurement of the expectation values
    in Eq.~\eqref{energy}. The two-qubit logical circuit is shown
    above the corresponding six-qubit encoded circuit. Qubits $a_1$
    and $a_2$ are ancillas. (a) The first section of the encoded
    circuit prepares the $\overline{\ket{00}}$ logical state. Ancilla
    $a_1$ is used to detect errors during the preparation. (b) The
    middle circuit sections apply the UCC exponential. We use ancilla
    $a_2$ to implement the rotation $\overline{R_y^1(\theta)}$. (c)
    The last circuit sections measure the expectation values. Gates
    $R_t$ perform a basis transformation that depends on the measured
    term.}
\end{figure*}

The VQE algorithm performs particularly well for this problem. It is a
hybrid quantum-classical algorithm that uses a quantum computer to
create and measure the properties of a parametrized trial wavefunction
and a classical computer to optimize the wavefunction parameters. Our
trial wavefunction is the unitary coupled-cluster (UCC)
ansatz~\cite{Bartlett:1989, Taube:2006}. Its realization on quantum
computers in context of quantum chemistry has been studied in
Ref. ~\cite{Peruzzo:2014, OMalley:2016, Hempel:2018} and in our case
is given by
\begin{equation}
  \label{ansatz}
  \ket{\psi(\theta)} = e^{-i \theta Y_1 X_2 / 2} \ket{\Phi},
\end{equation}
where $\theta$ is a parameter and $\ket{\Phi} = \ket{00}$ is the
Hartree--Fock wavefunction. The ansatz energy is given by
\begin{equation}
  \label{energy}
  E(\theta) = g_1 + g_2 \ev{Z_1}_\theta + g_3 \ev{Z_2}_\theta + g_4
  \ev{Z_1 Z_2}_\theta + g_5 \ev{X_1 X_2}_\theta,
\end{equation}
where $\ev{O}_\theta = \bra{\psi(\theta)} O \ket{\psi(\theta)}$. VQE
uses a quantum computer to estimate the expectation values included in
$E(\theta)$ and a classical optimizer to find the value of $\theta$
that minimizes $E(\theta)$. Since our ansatz depends on a single
parameter only, we sample the full domain of $\theta$ and use a
peak-finding routine to minimize $E(\theta)$. It is then sufficient to
sample the individual expectation values in Eq.~\eqref{energy} only
once and use the same data with any set of coefficients $g_j$. A
quantum circuit that implements VQE is shown in the top of
Fig.~\ref{circuits}.

\section{Error-detecting code}

Our goal is to compare the performance of a circuit implemented with
physical qubits to a circuit implemented with logical qubits of the
[[4, 2, 2]] code. This code maps two logical qubits into a subspace of
four physical qubits as
\begin{equation}
  \label{code}
  \begin{aligned}
    \overline{\ket{00}} & = \frac{1}{\sqrt{2}} \left( \ket{0000} +
    \ket{1111} \right), \\ \overline{\ket{01}} & = \frac{1}{\sqrt{2}}
    \left( \ket{0011} + \ket{1100} \right), \\ \overline{\ket{10}} & =
    \frac{1}{\sqrt{2}} \left( \ket{0101} + \ket{1010} \right),
    \\ \overline{\ket{11}} & = \frac{1}{\sqrt{2}} \left( \ket{0110} +
    \ket{1001} \right), \\
  \end{aligned}
\end{equation}
where an overline denotes a logical wavefunction. This mapping allows
for the detection of one single-qubit error. To implement the circuit,
we have to construct the required logical gates from the set of
available physical gates. Our set of physical gates is limited to
arbitrary single-qubit gates and $\mathit{CNOT}$ gates between any
pairs of physical qubits.

The encoded circuit is shown in the bottom of Fig.~\ref{circuits}. Its
first part is a preparation of the initial logical state
$\overline{\ket{00}}$. The circuit uses an ancilla measurement to
detect an error during the preparation~\cite{Gottesman:2016}. The
measurement outcome zero corresponds to no error while the outcome one
signals an error.

Some logical gates can be implemented easily because the corresponding
physical gates act transversally, i.e., they can be implemented with
only single-qubit physical gates. The [[4, 2, 2]] code also
facilitates a very simple implementation of the logical
$\mathit{CNOT}$ gates as $\overline{\mathit{CNOT}}_{12} =
\mathit{SWAP}_{12}$ and $\overline{CNOT}_{21} = \mathit{SWAP}_{13}$,
where an overline denotes a logical gate and $\mathit{SWAP}_{ij}$
swaps physical qubits $i$ and $j$~\cite{Harper:2019}. We implement
$\mathit{SWAP}_{ij}$ and therefore the $\overline{\mathit{CNOT}}$
gates without performing any physical operation by relabelling the
respective qubits.

The arbitrary-angle rotation of the first logical qubit
$\overline{R_y^1(\theta)}$ cannot be implemented transversally. We
apply this gate by entangling the logical qubit with an ancilla and
performing a rotation and a measurement on the ancilla. The
measurement projects the wavefunction onto a rotated logical
state. The rotation circuit applies a $\theta$-rotation and a
$-\theta$-rotation to the $\overline{\ket{0}}$ and
$\overline{\ket{1}}$ states of the first logical qubit,
respectively. The complete circuit performs a $\theta$-rotation
because our logical wavefunction is initially prepared in the
$\overline{\ket{00}}$ state. A general gate would require additional
physical gates. Measured value zero in the ancilla corresponds to a
rotation by $\theta$ while one corresponds to a rotation by $\theta +
\pi$. We use both outcomes to sample the Hamiltonian terms.

Qubits are measured in the computational basis. Expectation value
measurements require basis transformations that are performed with
gates $R_t$. In particular, $R_t = I$ for the $\ev{Z_1}_\theta$,
$\ev{Z_2}_\theta$, and $\ev{Z_1 Z_2}_\theta$ terms as the respective
operators are already diagonal in the computational basis, and $R_t =
H$ for the $\ev{X_1 X_2}_\theta$ term. We detect a single bit-flip or
a single phase-flip error by calculating the parity of the measured
code qubits.

Our encoded circuit is not fully fault-tolerant. In particular, not
all single-qubit errors in logical rotation $\overline{R_y^1(\theta)}$
can be detected. We can also detect a bit-flip or a phase-flip, but
not both at the same time (see Appendix~\ref{measurement} for
details).

\section{Experiment}

The algorithm can be summarized as follows. We sample the
$\ev{Z_1}_\theta$, $\ev{Z_2}_\theta$, $\ev{Z_1 Z_2}_\theta$, and
$\ev{X_1 X_2}_\theta$ terms for $\theta \in [-\pi, \pi)$ on a quantum
computer. The $\ev{Z_1}_\theta$, $\ev{Z_2}_\theta$, and $\ev{Z_1
  Z_2}_\theta$ terms are measured with a single circuit without any
basis transformations. We execute the circuit with $R_t = H$ to
measure the $\ev{X_1 X_2}_\theta$ term. The ground-state energy for
each internuclear separation $R$ is then calculated by minimizing
$E(\theta)$. Unlike classical variational algorithms, the minimal
energy can be lower than the exact energy due to systematic errors
and noise.

We ran both the two-qubit logical circuit and the six-qubit encoded
circuit on the Tokyo chip on IBM Q Experience. The major errors on
this platform are readout errors~\cite{Kandala:2017, Dumitrescu:2018,
  YeterAydeniz:2019}. If a qubit is in the $\ket{0}$ state, there is a
significant probability of measuring outcome one and vice versa. The
readout errors are asymmetric, i.e., the probability of measuring zero
when a qubit state is $\ket{1}$ is higher than the probability of
measuring one when the state is $\ket{0}$. This is mostly due to the
readout time being significant in comparison to the $T_1$ coherence
time, so the qubit can decay from the $\ket{1}$ state to the $\ket{0}$
state during the readout. We employed a readout error correction
technique known as unfolding based on a Bayesian probabilistic
model~\cite{Nachman:2019}. We first measured and estimated the
probability of each outcome when the qubits were prepared in each
computational basis state. We then used this probability matrix to
iteratively unfold all measured counts to corresponding true counts
(see Appendix~\ref{correction} for details).

\begin{figure*}
  \centering
  \includegraphics{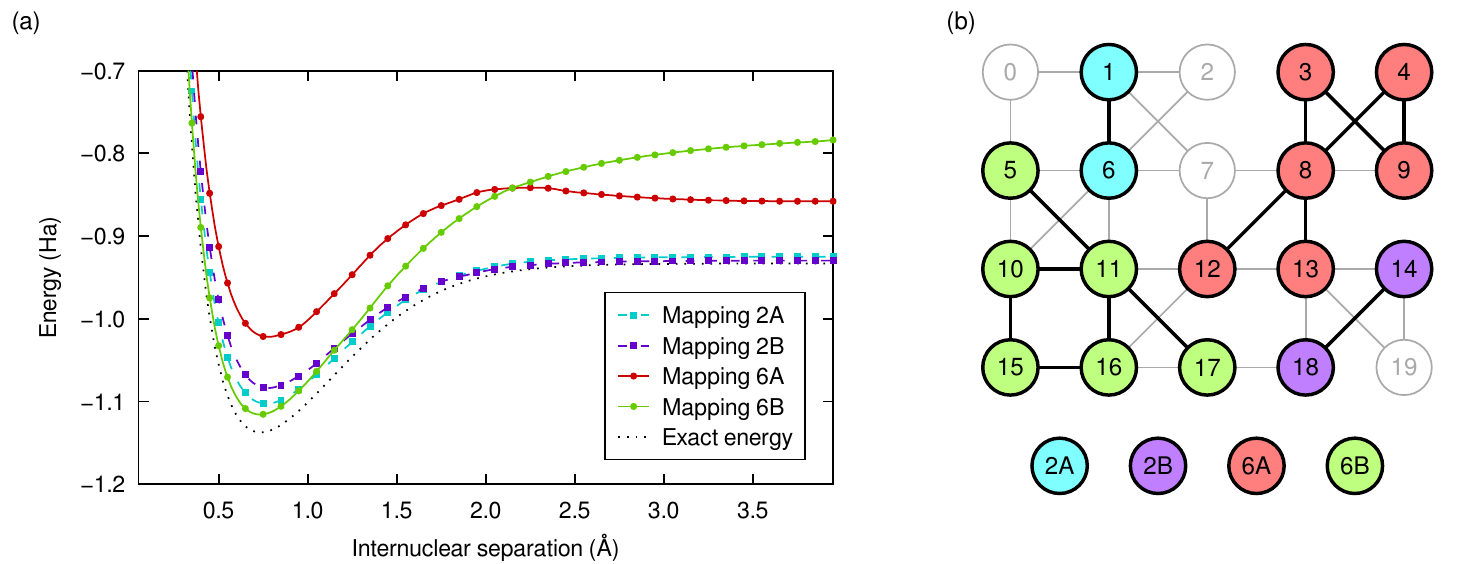}
  \caption{\label{qubits} Variability of results with qubit mapping.
    (a) Examples of potential energy curves obtained using two random
    mappings for each circuit on the 20-qubit Tokyo chip. (b) Chip
    geometry with highlighted two-qubit mappings 2A and 2B, and
    six-qubit mappings 6A and 6B.}
\end{figure*}

The chip contained 20 qubits arranged in a two-dimensional
geometry. There were 72 ways to map our two-qubit physical circuit and
288 ways to map our six-qubit encoded circuit to the chip qubits. We
found that the results depended significantly on the chosen qubits and
also on the order of the applied gates. The result variability is
illustrated in Fig.~\ref{qubits}, where 2A and 2B denote two-qubit
mappings $\left( q_1, q_2 \right) = \left( 1, 6 \right)$ and $\left(
q_1, q_2 \right) = \left( 14, 18 \right)$, and 6A and 6B denote
six-qubit mappings $\left( a_1, a_2, q_1, q_2, q_3, q_4 \right) =
\left( 13, 9, 8, 4, 3, 12 \right)$ and $\left( a_1, a_2, q_1, q_2,
q_3, q_4 \right) = \left( 5, 15, 11, 16, 10, 17 \right)$. The $\ev{X_1
  X_2}_\theta$ term is the most sensitive term in
Eq.~\eqref{energy}. To find an optimal mapping, we measured $\ev{X_1
  X_2}_\theta$ for $\theta = -3\pi/4$, $-\pi/2$, $-\pi/4$, $0$,
$\pi/4$, $\pi/2$, and $3\pi/4$, applied readout error correction, and
calculated the $L^1$ distances between the corrected results and the
exact results for each mapping. We used the mappings with the smallest
distances to run the final circuits. The compiler reordered gates
based on the qubit mapping, so this technique took into account both
the qubit mapping and the gate order variability.

\begin{figure*}
  \centering
  \includegraphics{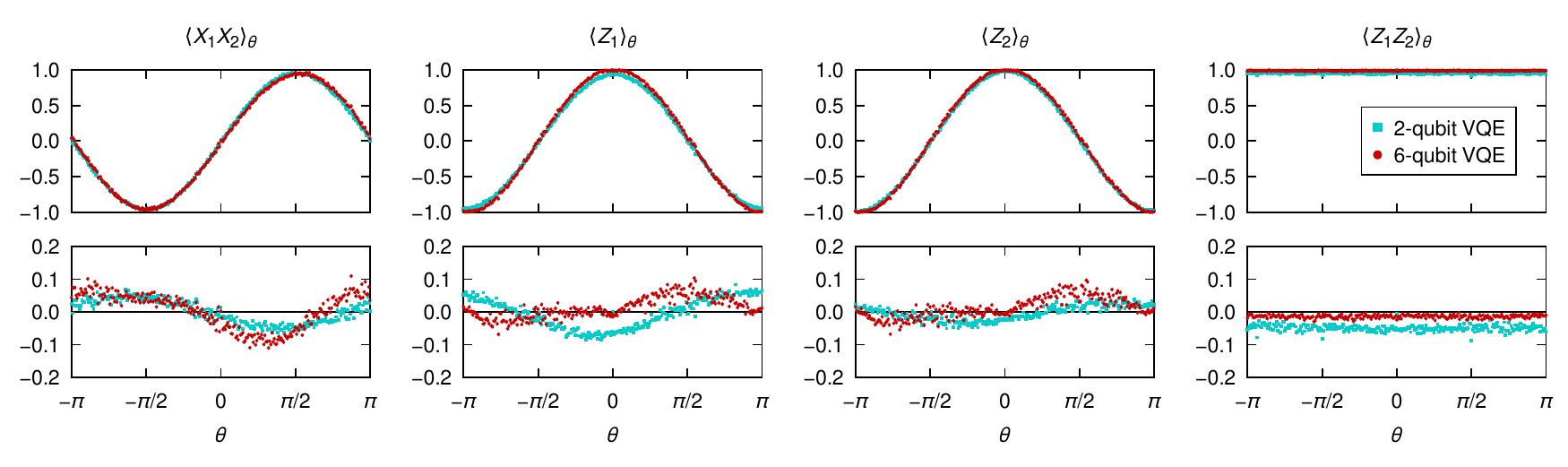}
  \caption{\label{terms} Measured expectation values of terms in the
    Hamiltonian for both the two-qubit and the six-qubit circuit. The
    top panels show the expectation values and the bottom panels show
    their differences from the exact values. The spread in values of
    neighboring points is due to shot noise.}
\end{figure*}

\begin{figure*}
  \centering
  \includegraphics{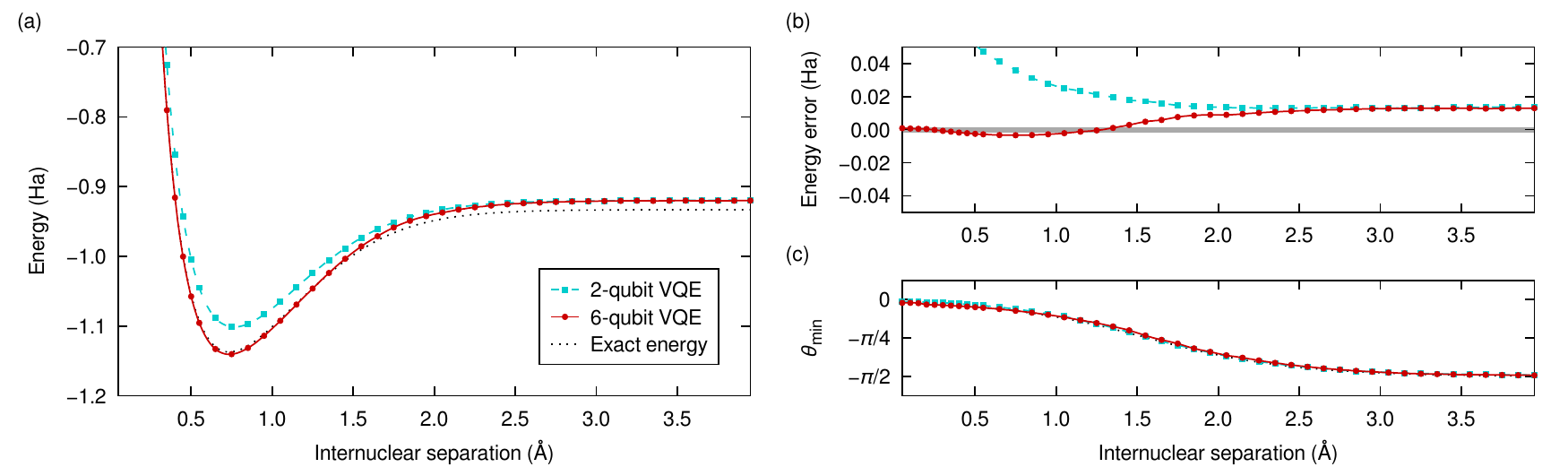}
  \caption{\label{experiment} Comparison of VQE results obtained with
    the two-qubit and the six-qubit circuit using the best available
    qubits. (a) Energy potential curve of the $\mathrm{H}_2$
    molecule. The exact energy is the lowest eigenvalue of
    Hamiltonian~\eqref{hamiltonian}. (b) Difference between the
    measured energy and the exact energy. Gray band shows the range of
    chemical accuracy ($1.6 \times 10^{-3} \, \mathrm{Ha}$). (c) Value
    of the UCC parameter $\theta$ at the energy minimum.}
\end{figure*}

We executed the final calculations for both the two-qubit and the
six-qubit circuit using the optimal mappings $\left( q_1, q_2 \right)
= \left( 13, 18 \right)$ for the two-qubit circuit and $\left( a_1,
a_2, q_1, q_2, q_3, q_4 \right) = \left( 12, 5, 11, 6, 10, 17 \right)$
for the six-qubit circuit. The $\ev{Z_1}_\theta$, $\ev{Z_2}_\theta$,
$\ev{Z_1 Z_2}_\theta$, and $\ev{X_1 X_2}_\theta$ terms were obtained
for 257 values of $\theta$ in the $[-\pi, \pi]$ interval. Each
measured value was sampled with 8192 shots. For the encoded six-qubit
circuit we postselected the outcomes based on their ancilla values. In
particular, we measured all six qubits, performed readout error
correction, and discarded outcomes with value one in the first ancilla
and outcomes outside of the code space. Outcomes with values zero and
one in the second ancilla were processed as samples corresponding to
rotations by $\theta$ and $\theta \pm \pi$, respectively. We summed
the renormalized counts of constituent basis states in
Eq.~\eqref{code} to calculate the logical state counts. The calculated
expectation values of the Hamiltonian terms are shown in
Fig.~\ref{terms}. We then used a peak-finding routine to find $\theta$
that minimized the energy in Eq.~\eqref{energy} for each internuclear
separation. The calculated energy potential curves are shown in
Fig.~\ref{experiment}. The results demonstrate that the six-qubit
encoded circuit improves the accuracy of the ground-state energy. The
$\ev{Z_1}_\theta$ and $\ev{Z_2}_\theta$ terms contribute to
$E(\theta)$ the most at small internuclear separations $R$ while the
$\ev{X_1 X_2}_\theta$ term is dominant at large $R$. The encoded
circuit performs better especially at small $R$ where $\theta \approx
0$. Both energy potential curves are slightly inaccurate at large
$R$. These inaccuracies can be fully explained by errors in $\ev{X_1
  X_2}_\theta$ at $\theta \approx -\pi/2$. The $\ev{X_1 X_2}_\theta$
term is very sensitive to the quality of Hadamard gates applied before
the measurement. Small inaccuracies lead to many nonvanishing
coefficients in the final wavefunction. Our results suggest that the
[[4, 2, 2]] code lacks the power to reliably detect errors in the
$\ev{X_1 X_2}_\theta$ term on this hardware (see
Appendix~\ref{measurement} for details).

\section{Discussion}

Our encoded circuit requires more physical qubits and gates than our
logical circuit and is therefore more sensitive to errors. However,
the gain by using the code was larger than the loss due to the circuit
complexity. The results show that quantum error detection is already
useful on NISQ devices even without achieving full fault
tolerance. The presented method can be used in addition to other error
mitigation techniques. Our implementation uses two ancillary qubits
with postselection on their measured outcomes. In principle, it would
be possible to use just one ancilla if we had an ability to perform a
qubit reset. Similarly, the postselection in the rotation gate would
be unnecessary if we had an ability to apply conditional gates
dependent on measurement outcomes.

Some of the previous VQE experiments~\cite{OMalley:2016, Kandala:2017,
  Colless:2018, Hempel:2018, Kandala:2019, Ganzhorn:2019} found the
ground-state energy of the $\mathrm{H}_2$ molecule with a comparable
or better accuracy. They used techniques like higher qubit states
measurement~\cite{OMalley:2016}, quantum subspace
expansion~\cite{Colless:2018}, and noise
extrapolation~\cite{Kandala:2019} to mitigate errors. We emphasize
that our circuits do not use any such techniques. Our QEC method
demonstrates that on the same hardware and using the same algorithm,
the encoded circuit results in smaller errors than the physical
circuit. Other error mitigation techniques are complementary to the
presented method.

\begin{acknowledgments}

We thank Jarrod R. McClean, Mekena Metcalf, and Shaobo Zhang for
helpful comments. This work was supported by the DOE under contract
DE-AC02-05CH11231, through the Office of Advanced Scientific Computing
Research (ASCR) Quantum Algorithms Team Program, and the Office of
High Energy Physics through the Quantum Information Science Enabled
Discovery (QuantISED) program (KA2401032). This research used
resources of the Oak Ridge Leadership Computing Facility, which is a
DOE Office of Science User Facility supported under Contract
DE-AC05-00OR22725.

\end{acknowledgments}

\appendix

\section{Hamiltonian transformation}

We use a transformation presented in Ref.~\cite{Colless:2018} to map
the electronic-structure space to qubits. The transformed space
corresponds to a $\mathrm{H}_2$ molecule with two electrons and zero
total spin. In particular,
\begin{equation}
  \begin{aligned}
    a_{1\uparrow}^\dagger a_{1\downarrow}^\dagger \ket{\mathrm{vac}} &
    \to \ket{00}, \\ a_{1\uparrow}^\dagger a_{2\downarrow}^\dagger
    \ket{\mathrm{vac}} & \to \ket{01}, \\ a_{2\uparrow}^\dagger
    a_{1\downarrow}^\dagger \ket{\mathrm{vac}} & \to \ket{10},
    \\ a_{2\uparrow}^\dagger a_{2\downarrow}^\dagger
    \ket{\mathrm{vac}} & \to \ket{11}, \\
  \end{aligned}
\end{equation}
where $a_{is}^\dagger$ is an operator that creates an electron with
spin $s$ in orbital $i$ and $\ket{\mathrm{vac}}$ is the vacuum state.

\section{Analytical model}

\begin{figure}
  \centering
  \includegraphics{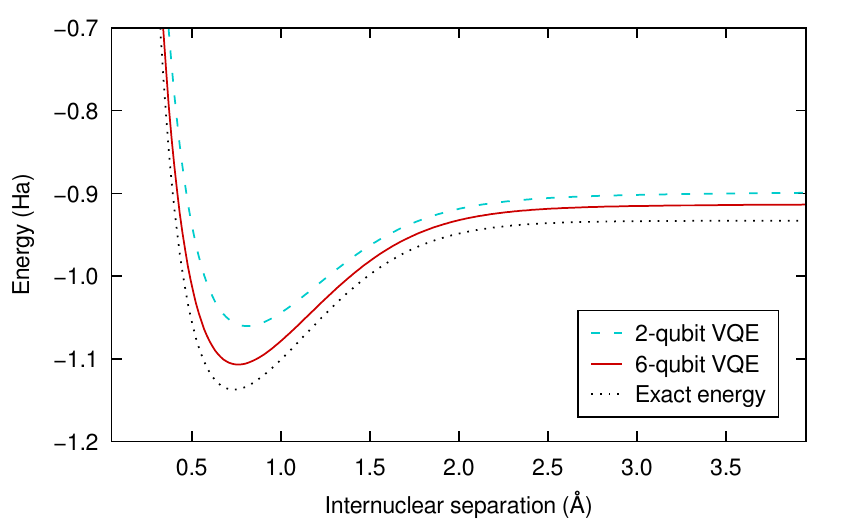}
  \caption{\label{model} Energy potential curves of the $\mathrm{H}_2$
    molecule calculated analytically using the VQE algorithm with the
    depolarizing noise model. The two-qubit gate error rate is $p =
    5\,\%$.}
\end{figure}

\begin{figure}
  \centering
  \includegraphics{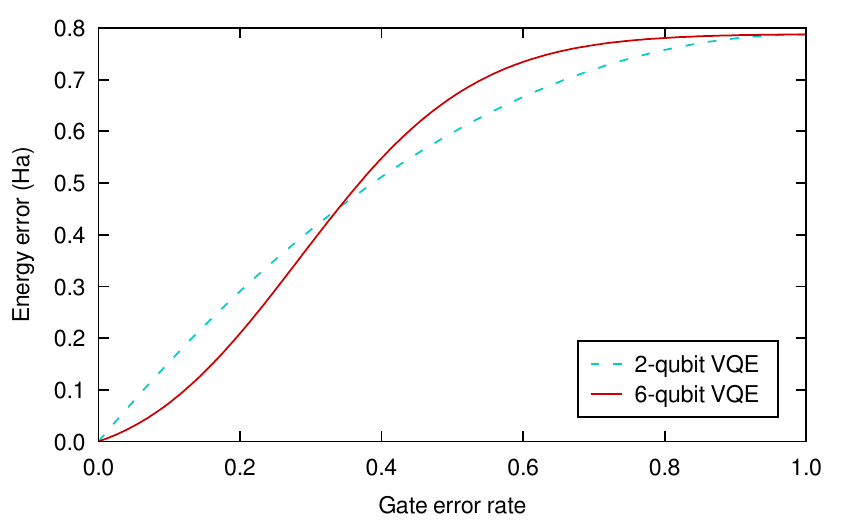}
  \caption{\label{noise} Energy error with the depolarizing noise
    model for internuclear separation $R = 0.75\,\text{\AA}$. The
    six-qubit encoded circuit performs better for error rates up to
    about $30\,\%$.}
\end{figure}

We analyze the effect of noise on the calculated ground-state energies
using the depolarizing noise model. The noise operation for one qubit
is given by~\cite{Nielsen:2010}
\begin{equation}
  \epsilon(\rho) = (1-p) \rho + p \frac{I}{2},
\end{equation}
where $\rho$ is the density matrix and $p$ is the probabilistic error
rate. The value of $p = 0$ corresponds to vanishing noise and $p = 1$
corresponds to full noise. We assume that the noise affects only
qubits involved in a particular gate application. Separate operations
are used for one-qubit gates,
\begin{equation}
  \epsilon_i(\rho) = (1 - p_1) \rho + \frac{p_1}{4} \sum_{E_i \in
    \mathcal{P}_i} E_i^\dagger \rho E_i,
\end{equation}
and for two-qubit gates,
\begin{equation}
  \epsilon_{i, j} (\rho) = (1 - p_2) \rho + \frac{p_2}{16}
  \sum_{\substack{E_i \in \mathcal{P}_i \\ E_j \in \mathcal{P}_j}}
  E_i^\dagger E_j^\dagger \rho E_i E_j,
\end{equation}
where $\mathcal{P}_i = \{I_i, X_i, Y_i, Z_i\}$ is the set of the unit
matrix and the Pauli matrices acting on qubit $i$. The noise
operations above are performed on the density matrix after each gate
application to a respective set of qubits. We characterize the noise
channel with only a single parameter $p$ and use $p_2 = p$ and $p_1 =
p/16$ since the single-qubit gates have significantly higher
fidelities in hardware. The comparison of the ground-state energy
calculated with the noisy physical and encoded circuits is shown in
Fig.~\ref{model}. The energy error as a function of $p$ for a selected
internuclear separation is shown in Fig.~\ref{noise}. The results show
that the encoded circuit outperforms the physical circuit when the
error probability of two-qubit gates is less than about $30\,\%$. This
threshold is significantly higher than the error rates for two-qubit
gates on the Tokyo chip which are less than $5\,\%$. Assuming that the
depolarization noise model is an appropriate error model, the encoded
circuit should produce a better energy estimate.

\section{\label{correction} Readout error correction}

Correcting measurements of discrete data for readout bias has a long
history. For example, in high energy physics experiments, binned
differential cross sections are corrected for detector effects in
order to compare them with predictions from quantum field theory. In
that context, the corrections are called unfolding (sometimes called
deconvolution in other fields) and a variety of techniques have been
proposed and are in active use~\cite{Cowan:2002, Blobel:2013}. Quantum
readout error correction can be represented as a binned unfolding
where each bin corresponds to one of the possible $2^n$
configurations, where $n$ is the number of qubits.

\begin{figure}
  \centering
  \includegraphics{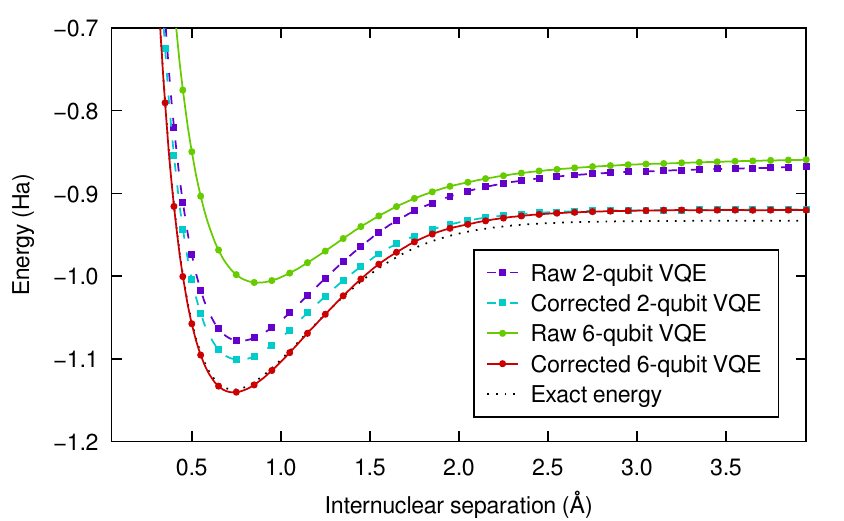}
  \caption{\label{raw} Comparison of potential energy curves obtained
    with raw measurement outcomes and with outcomes corrected for
    readout errors for both the two-qubit and six-qubit circuits.}
\end{figure}

We use an iterative Bayesian unfolding technique~\cite{Lucy:1974,
  Richardson:1972, DAgostini:1995}. Given a response matrix
\begin{equation}
  R_{ij} = \Pr(\text{measure}\ i \,|\, \text{truth is}\ j),
\end{equation}
a measured spectrum $m_i = \Pr(\text{measure}\ i)$ and a prior truth
spectrum $t_i^0 = \Pr(\text{truth is}\ i)$, the iterative technique
proceeds according to an equation
\begin{equation}
  \label{unfolding}
  \begin{split}
    t_i^{l + 1} & = \sum_j \Pr(\text{truth is}\ i \,|\,
    \text{measure}\ j) \times m_j \\ & = \sum_j \frac{R_{ji}
      t_i^l}{\sum_k R_{jk} t_k^l} \times m_j,
  \end{split}
\end{equation}
where $l$ is the iteration number. The advantage of
Eq.~\eqref{unfolding} over simple matrix inversion is that the result
is a probability (nonnegative and unit measure). We construct $R_{ij}$
by preparing $2^n$ calibration circuits where each qubit computational
state is constructed with $X$ gates. The entries of $R_{ij}$ are the
fraction of measurements that qubit configuration $i$ is observed in
configuration $j$. We use a uniform distribution as the initial
spectrum $t_i^0$. The iterative procedure described in
Eq.~\eqref{unfolding} is repeated until convergence. The effect of
readout error correction on potential energy curves is shown in
Fig.~\ref{raw}. Comparison of iterative Bayesian unfolding with other
available methods is discussed in Ref.~\cite{Nachman:2019}.

\section{Initial state preparation}

\begin{figure}
  \centering
  \includegraphics{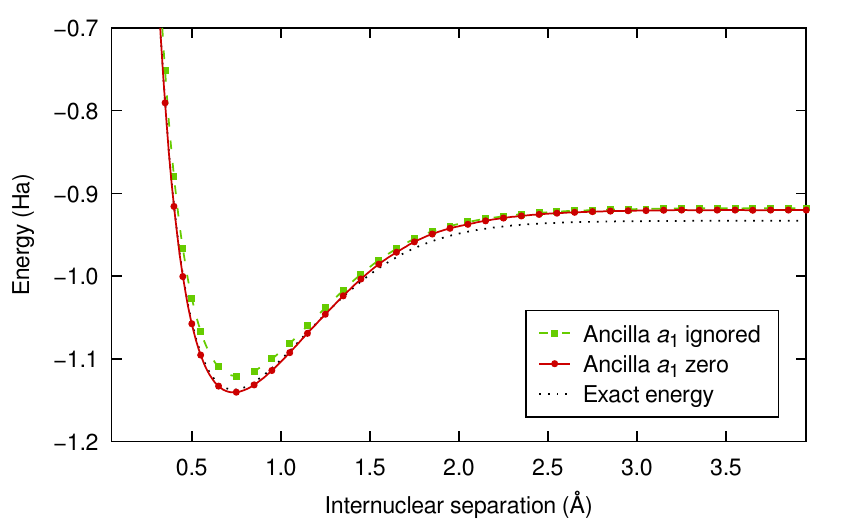}
  \caption{\label{ancilla} Comparison of potential energy curves
    obtained with the six-qubit circuit when ignoring the value of
    ancilla $a_1$ and when postselecting only outcomes with $a_1$
    being zero.}
\end{figure}

The encoded circuit uses ancilla $a_1$ to detect an error during the
initial state preparation. Measured value zero corresponds to no error
whereas one corresponds to a detected error in the state
preparation. We therefore postselect only outcomes with $a_1$ being
zero. Fig.~\ref{ancilla} shows the effect of the postselection. About
8\% of the samples were discarded due to the postselection on $a_1$.

\section{\label{measurement} Syndrome measurement}

The stabilizers of the [[4, 2, 2]] code are generated by operators
$X_1 X_2 X_3 X_4$ and $Z_1 Z_2 Z_3 Z_4$~\cite{Gottesman:2016}. Code
words are eigenvectors of stabilizers with $+1$ eigenvalue. A single
bit-flip error in a code word transforms the code word into an
eigenvector of the $Z_1 Z_2 Z_3 Z_4$ stabilizer with $-1$
eigenvalue. A single phase-flip error transforms it into an
eigenvector of the $X_1 X_2 X_3 X_4$ stabilizer with $-1$
eigenvalue. To detect both a bit-flip and a phase-flip error, it is
necessary to perform syndrome measurements in the two bases
corresponding to the two stabilizer generators~\cite{Devitt:2013}. We
do not perform these measurements in our encoded circuit. Instead, we
only measure physical qubits in the computational basis. The parity of
the code qubits then corresponds to the eigenvalue of one of the
generators. In particular, it corresponds to $Z_1 Z_2 Z_3 Z_4$ when
$R_t = I$ and to $X_1 X_2 X_3 X_4$ when $R_t = H$. We therefore
perform only one syndrome measurement and detect only a bit-flip or a
phase-flip error. A detection of both errors would require additional
qubits and gates~\cite{Devitt:2013} or additional
measurements~\cite{McClean:2020}. About 11\% and 16\% of samples were
discarded due to syndrome measurement when $R_t = I$ and $R_t = H$,
respectively, for data used in the final figures. The error rate for
the $\ev{X_1 X_2}_\theta$ term is therefore significantly higher than
for the other Hamiltonian terms.

\section{Qubit mappings}

The availability of qubits and their connections has changed during
the data collection. The final data were collected after a connection
between qubits three and nine was turned off. Additionally, qubit
seven was not available during experiments with the six-qubit
circuit. As a result, there were only 70 and 116 possible mappings
from the abstract qubits to the physical qubits for the two-qubit and
the six-qubit circuits, respectively.

We found it practical to run our circuits for each of their possible
mappings to find optimal mappings. However, this approach is
unfeasible for larger systems. An alternative method is to estimate
the total circuit fidelity from reported gate fidelities. Although we
found acceptable qubit mappings using this approach, we were never
able to find the best one. The action of gates is highly nontrivial
and cannot be reduced to a single number. A significant error source
is also cross-talk between qubits. Estimates of the total circuit
fidelities therefore have only a limited reliability. Optimal qubit
mapping is an area of active research~\cite{Murali:2019}.

\bibliographystyle{apsrev4-1}
\bibliography{main}

\end{document}